\documentstyle[epj,epsfig,final]{svjour}

\begin{document}
\title{Calculation of photoemission spectra of the doped Mott
insulator La$_{1-x}$Sr$_{x}$TiO$_{3}$ using LDA+DMFT(QMC)}
\author{I.A.~Nekrasov$^{1,}$\thanks{\email{nekrasov@ifmlrs.uran.ru}}, K. Held$^{2,}$\thanks{\email{held@physik.uni-augsburg.de}}, N. Bl\"{u}mer$^{2}$, A.I.~Poteryaev$^{1}$,
 V.I.~Anisimov$^{1}$,
and D. Vollhardt$^{2}$}
\authorrunning{I.A. Nekrasov et al.}
\titlerunning{Calculation of photoemission spectra of
La$_{1-x}$Sr$_{x}$TiO$_{3}$ using LDA+DMFT(QMC)}
\institute{$^{1}$Institute of Metal Physics, Russian Academy of Sciences-Ural Division,%
\\
620219 Yekaterinburg GSP-170, Russia\\
$^{2}$Theoretische Physik III, Elektronische Korrelationen und Magnetismus,\\
Universit{\"{a}}t Augsburg, D-86135 Augsburg, Germany}
\date{\today}

\abstract{
The spectral properties of La$_{1-x}$Sr$_{x}$TiO$_{3}$, a doped Mott
insulator with strong Coulomb correlations, are calculated with the {\em %
ab initio} computational scheme LDA+DMFT(QMC). It starts from the
non-interacting electronic band structure as calculated by the local density
approximation (LDA), and introduces the missing correlations by the
dynamical mean-field theory (DMFT), using numerically exact quantum
Monte-Carlo (QMC) techniques to solve the resulting self-consistent
multi-band single-impurity problem. The results of the LDA+DMFT(QMC)
approach for the photoemission spectra of La$_{1-x}$Sr$_{x}$TiO$_{3}$ are in
good agreement with experiment and represent a considerable qualitative and
quantitative improvement on standard LDA calculations.
}

\maketitle

\section{Introduction}

At present, the electronic properties of solids are investigated by two
essentially separate communities, one using model Hamiltonians in
conjunction with many-body techniques, the other employing density
functional theory (DFT)~\cite{Hohenberg64}. DFT and its local density
approximation (LDA) have the advantage of being {\em ab initio} approaches
which do not require empirical parameters as input. Indeed, they are highly
successful techniques for the calculation of the electronic structure of
real materials~\cite{JonesGunn}. However, in practice DFT/LDA is seriously
restricted in its ability to describe strongly correlated materials where
the on-site Coulomb interaction is comparable with the band width. Here, the
model Hamiltonian approach is more general and powerful since there exist
systematic theoretical techniques to investigate the many-electron problem
with increasing accuracy. Nevertheless, the uncertainty in the choice of the
model parameters and the technical complexity of the correlation problem
itself prevent the model Hamiltonian approach from being
a flexible or reliable
enough tool for studying real materials. The two approaches are therefore
complementary. In view of the individual power of DFT/LDA and the
model Hamiltonian approach, respectively, it would be highly desirable to be
able to combine these techniques, thereby creating an enormous potential for
future {\em ab initio} investigations of all real materials, including,
e.g., $f$-electron systems and Mott insulators. One of the first successful
attempts in this direction was the LDA+U method~\cite{Anisimov91}, which
combines LDA with a basically static, i.e., Hartree-Fock-like, mean-field
approximation for a multi-band Anderson lattice model
(with interacting and non-interacting orbitals). This method proved to
be a very useful tool in the study of long-range ordered, insulating states
of transition metals and rare-earth compounds. However, the paramagnetic
metallic phase of correlated electron systems such as high-temperature
superconductors and heavy-fermion systems clearly requires a treatment that
goes beyond a static mean-field approximation and includes dynamical
effects, e.g., the frequency dependence of the self-energy.

During the last decade a new many-body approach was developed which is
especially well-suited for the investigation of strongly correlated metals
-- the dynamical mean-field theory (DMFT)~\cite{vollha93,pruschke,georges96}.
It becomes exact in the limit of high lattice coordination numbers~\cite
{MetzVoll89} and preserves the dynamics of local interactions; hence it
represents a {\em dynamic} mean-field approximation. In this
non-perturbative approach, the lattice problem is mapped onto an effective
Anderson impurity model with a hybridization function which has to be
determined self-consistently. To solve the effective impurity problem one
can either use approximative techniques such as iterated perturbation
theory (IPT)~\cite{georges92,georges96} and the non-crossing approximation
(NCA) \cite{NCA1,pruschke89,NCA2}, or employ numerical techniques like
quantum Monte-Carlo simulations (QMC)~\cite{QMC}, exact diagonalization
(ED)~\cite{caffarel94,georges96}, and numerical renormalization group (NRG)~\cite{NRG}.
In principle, QMC, ED and NRG are exact methods, but
require an extrapolation: discretization of the imaginary time $\Delta \tau
\rightarrow 0$ (QMC), the number of lattice sites of the respective impurity model $%
n_{s}\rightarrow \infty $ (ED), and the parameter for logarithmic discretization of the
conducting band $\Lambda \rightarrow 1$ (NRG), respectively. 

In principle, the main idea of the LDA+U method~\cite{Anisimov91}
(i.e., complementing the LDA band structure by a screened
Coulomb interaction between localized tight-binding orbitals)
can be practically applied with more refined approximation schemes~\cite
{poter97,lichten98,janis98,laegsgaard,wolenski98,zoelfl99}. Indeed, a
calculation scheme supplementing LDA with DMFT to include dynamic effects
was first formulated by Anisimov et al.~\cite{poter97} and was used to
calculate the photoemission spectra of La$_{1-x}$Sr$_{x}$TiO$_{3}$, a doped
Mott-insulator and strongly correlated paramagnetic metal, in connection
with IPT~\cite{poter97} and NCA~\cite{zoelfl99}. The LDA++ approach by
Lichtenstein and Katsnelson~\cite{lichten98} formulates a very similar
strategy and was recently applied to investigate correlation effects in iron~\cite{Kats98,kats99}.
Both IPT and NCA are approximative methods to solve the
effective single-impurity problem in the LDA+DMFT scheme. They have the
advantage of being numerically inexpensive, but their reliability,
especially in the case of multi-band systems with particle densities off
half-filling, is in principle uncertain. In this situation, it is clearly
desirable to employ a controlled computation scheme to obtain numerically
exact results from LDA+DMFT. The QMC method is such a scheme and was already
applied by Lichtenstein and collaborators to calculate the magnetic excitation spectrum of
ferromagnetic iron~\cite{kats99} and the photoemission spectra of
Sr$_{2}$RuO$_{4}$~\cite{liebsch00}.
In the present paper we report on our implementation of the LDA+DMFT(QMC)
technique and the results obtained with it for the photoemission spectra of~La$_{1-x}$Sr$_{x}$TiO$_{3}$.

\section{Computational scheme}

\subsection{Local density approximation (LDA)}

The main problem in combining LDA and model Hamiltonian approaches comes
from the fact that their foundations are very different. The LDA employs a
functional of the electron density, while Hubbard and Anderson models are
formulated in terms of localized, site-centered, atomic-like orbitals. In
order to merge the approaches it is necessary to write LDA equations in the
basis of such orbitals. There is one variant of the LDA method, based on
tight-binding linearized muffin-tin orbitals in the orthogonal approximation
(TBLMTO)~\cite{LMTO}, which is naturally realized in such a basis. The
corresponding Hamiltonian can be written as 
\begin{eqnarray}
H_{{\rm LDA}}=\sum_{ilm,{\rm }jl^{\prime }m^{\prime },\sigma }&&(\delta
_{ilm,jl^{\prime }m^{\prime }}\varepsilon _{ilm}{\hat{n}}_{ilm}^{\sigma
} \nonumber\\ 
&&+ t_{ilm,jl^{\prime }m^{\prime }}{\hat{c}}_{ilm}^{\sigma \dagger }{\hat{c}}%
_{jl^{\prime }m^{\prime }}^{\sigma }),  \label{hlmto}
\end{eqnarray}
where $i,j$ are site indices, $l,m,l^{\prime },m^{\prime }$ are orbital
indices, $\sigma $ is a spin index, and operators carry a hat.

Taking this expression as the non-interacting part of a multi-band periodic
Anderson model, one may complement it by a correlation term describing the
local contributions to the Coulomb interaction 
\begin{equation}
H_{{\rm corr}}=\quad \frac{1}{2}\sum_{il,m\sigma m^{\prime }\sigma
^{\prime }}\!\!\!\!\!\!\!\!\!\!\!\!\!\phantom{\sum_{m}}^{\!\!\!\!\!\prime
}\;\;U_{mm^{\prime }}^{il}\,\hat{n}_{ilm\sigma }\hat{n}_{ilm^{\prime }\sigma
^{\prime }}.  \label{Hcorr}
\end{equation}
The prime on the sum indicates that at least two of the indices on different
operators have to be different. Here, $U_{mm^{\prime }}^{il}$ denotes the
direct Coulomb integral. The much smaller exchange integral and other
local contributions of the  Coulomb interaction have been neglected. 
Furthermore, non-local Coulomb contributions are not considered
in $H_{\rm corr}$. Note, that the largest non-local contributions is the
nearest-neighbor density-density interaction which, to leading order in 
$Z$ ($Z$: number of nearest-neighbor sites), yields only the Hartree
term  \cite{MH1} which is already taken into account in the LDA.
In an actual calculation, it is
not possible to include the local two-particle interaction terms between all
orbitals appearing in (\ref{Hcorr}) since the
number of states grows exponentially with increasing orbital quantum number.
Thus, one usually
 concentrates on a certain subset of correlations and treats the
influence of the remaining states via the use of effective, screened
interaction parameters for the shells under consideration. In this spirit, we
will assume in the following that it is only necessary to take the Coulomb
interaction for the $d$-shell of the transition metal ions ($i=i_{d}$ and $%
l=l_{d}$) explicitly into account;
therefore the indices $il$ will be omitted. The correlation part of the
Hamiltonian then acts only on the $d$-wave functions of the transition metal
ions. All other valence orbitals will be treated as bands of itinerant
electrons which are well described by the LDA.

One must take into account, however, that the Coulomb interaction is already
present in LDA in  some averaged way. Hence, to avoid double-counting one
needs to subtract this term from the LDA-Hamiltonian. Unfortunately, there
exists no direct microscopic or diagrammatic link between the Hubbard model
approach and LDA, and it is thus not possible to express the LDA energy
rigorously via the $d-d$ Coulomb interaction parameter $U$ (except for the
atomic limit where one can make a connection between the
 Coulomb parameter $U$ of the Hubbard model and the second
derivative of the atomic total energy as a function of the number
of electrons). While it is known
that LDA eigenvalues are rather bad approximations for excitation energies
of systems with strong Coulomb interactions, the LDA {\em total} energy as a
function of the number of electrons is a much better approximation for the
exact functional. Furthermore, the values of $U$ obtained from LDA
calculations often agree well with experimental data and more rigorous
calculations. 

Therefore one may expect that the LDA part of the Coulomb interaction energy
is well approximated by the averaged value of the Coulomb interaction energy
in (2) 
\begin{equation}
 E_{{\rm corr}}{\large :}=\frac{1}{2}Un_{d}(n_{d}-1){\large .}
\label{ELDAalimit}
\end{equation}
Here, $U$ is the mean value of the Coulomb interaction and may be obtained
from a first-principles constrained LDA calculation~\cite{Parameters} or from experiment,
e.g. high-energy spectroscopy; $n_{d}$ is the total number of $d$-electrons. 

In LDA, one-electron energies are defined as derivatives of the total energy
as a function of the occupation numbers for the corresponding states. Hence,
the one-electron energy level for the {\em non-interacting} d-states is
obtained by~\cite{poter97} 
\begin{equation}
\varepsilon _{d}^{0}:=\frac{d}{dn_{d}}(E_{\rm LDA}-E_{{\rm corr}%
})=\varepsilon _{d}^{\rm LDA}-U(n_{d}-\frac{1}{2})
\end{equation}
with $\varepsilon _{d}^{\rm LDA}:=\frac{d}{dn_{d}}E_{\rm LDA}$, and $E_{\rm LDA}$ being the
total energy calculated from $H_{\rm LDA}$.

Then, the new non-interacting Hamiltonian will have the form 
\begin{eqnarray*}
H^{0}_{{\rm LDA}}\!&=&\!\!\!\sum_{ilm,jl^{\prime }m^{\prime },\sigma }\!\!\!(\delta _{ilm,jl^{\prime
}m^{\prime }}\varepsilon _{ilm}^{0}{\hat{n}}_{ilm}^{\sigma
}+t_{ilm,jl^{\prime }m^{\prime }}{\hat{c}}_{ilm}^{\sigma \dagger }{\hat{c}}%
_{jl^{\prime }m^{\prime }}^{\sigma }), 
 \end{eqnarray*}
where $\varepsilon
_{ilm}^{0}=\varepsilon _{ilm}\left( 1-\delta _{l,l_{d}}\right) +\varepsilon
_{d}^{0}\delta _{l,l_{d}}$. In
reciprocal space, the matrix elements of the operator $H^{0}_{{\rm LDA}}$ are given by: 
\begin{eqnarray}
(H^{0}_{{\rm LDA}}({\bf k}))_{qlm,q^{\prime }l^{\prime }m^{\prime }}&=&(H_{{\rm LDA}}({\bf k}))_{qlm,q^{\prime}l^{\prime }m^{\prime }}\nonumber \\
&&\!\!\! -\delta _{qlm,q^{\prime }l^{\prime}m^{\prime }}\delta _{ql,q_{d}l_{d}}U(n_{d}-\frac{1}{2}).
\end{eqnarray}
Here $q$ is an index of the atom in the elementary unit cell,
$(H_{{\rm LDA}}({\bf k}))_{qlm,q^{\prime}l^{\prime }m^{\prime }}$ is 
the matrix element in k-space of $H_{\rm LDA}$, 
and $q_{d}$ denotes the d-atoms in the unit cell.
This
non-interacting part $H^{0}_{{\rm LDA}}$, taken together with the interaction part (\ref
{Hcorr}),

\begin{equation}
 H=H^{0}_{{\rm LDA}}+H_{{\rm corr}}
\end{equation}
forms the {\em ab initio} Hamiltonian $H$ for a particular material
under investigation.

\subsection{Dynamical mean-field theory (DMFT)}

In general, the investigation of the correlated-electron Hamiltonian ${\cal H}
$ is too complicated to allow for an exact solution or even a numerical
investigation with more than about 10 sites. Here, the DMFT~\cite
{vollha93,pruschke,georges96} is a powerful approximation scheme which takes
into account electronic correlations and, in particular, correctly describes
the formation of a coherent quasiparticle band and incoherent Hubbard bands.

The DMFT maps the lattice problem onto a single-site problem, which is
equivalent to a multi-band single-impurity Anderson model, with the
self-consistency condition~\cite{poter97} (the $k$-integrated Dyson equation) 
\begin{eqnarray}
G_{qlm,q^{\prime }l^{\prime }m^{\prime }}(z)\!=\!\int\! \frac{{\rm d}{\bf k}}{V_{B}}&&
[z\delta_{qlm,q^{\prime }l^{\prime }m^{\prime }}-(H^{0}_{{\rm LDA}}({\bf k}))_{qlm,q^{\prime }l^{\prime }m^{\prime }}
\nonumber \\&& -
\delta _{ql,q_{d}l_{d}}
\Sigma_{qlm,q^{\prime }l^{\prime }m^{\prime }}(z)]^{-1}.
\end{eqnarray}
Here, $[...]^{-1}$ implies the inversion of the matrix with elements $n$ (=$%
qlm$), $n^{\prime }$(=$q^{\prime }l^{\prime }m^{\prime }$), 
and integration extends over the Brillouin zone with volume $%
V_{B}$. In the present study we consider a cubic-crystal
structure and assume the $t_{2g}$ orbitals to be interacting.
Due to the high symmetry of the crystal, these three $t_{2g}$ orbitals 
are degenerate. Without symmetry breaking, the Green function
and the self-energy remain degenerate,
i.e., $G_{qlm,q^{\prime }l^{\prime }m^{\prime }}(z)=G(z)\delta
_{qlm,q^{\prime }l^{\prime }m^{\prime }}$
and
$\Sigma_{qlm,q^{\prime }l^{\prime }m^{\prime }}(z)=\Sigma(z)\delta
_{qlm,q^{\prime }l^{\prime }m^{\prime }}$
for $l=l_d$ and $q=q_d$ (where $l_d$ and $q_d$ denote the Ti 
$t_{2g}$ orbitals).
If the partially filled band under consideration
is well separated from other bands, as in the case of 
La$_{1-x}$Sr$_{x}$TiO$_{3}$ and other transition metal oxides, one can describe
 the physics of  the partially filled band at the Fermi energy  approximately 
by an effective 
three band Hamiltonian $H^{0 \; {\rm eff}}_{{\rm LDA}}$, e.g., by downfolding 
to a basis with
$t_{2g}$ orbitals only. One  obtains  (indices $l=l_d$ and $q=q_d$ suppressed):
\begin{eqnarray}
\!\!\!\!\!\!G_{m m'}(z) &\!=\!&\!\int \! \frac{{\rm d} \bf k}{V_{B}}
[(z\!-\!\Sigma(z))\delta_{m,m^{\prime}}\!-\!(H^{0 \; \rm eff}_{{\rm LDA}}({\bf k}))_{m,m^{\prime }}
]^{-1}.
\end{eqnarray}
 Due to the diagonal structure of the self-energy
the interacting Green function can be expressed via the
 non-interacting Green function $G^{0}(z)$:
\begin{eqnarray}
G(z)&\!=\!&G^{0}(z-\Sigma (z))=\int d\omega \frac{\rho ^{0}(\omega )}{z-\Sigma
(z)-\omega },  \label{intg}
\end{eqnarray}
Thus,
it is possible to use the Hilbert transformation of the unperturbed
LDA-calculated density of states $\rho ^{0}(\omega )$ and we do so
in the following. This approximation is
justified if  the hybridization between the $t_{2g}$ orbitals and 
the other orbitals is rather weak as in the case of LaTiO$_3$.

The DMFT 
single-site problem depends on ${\cal G}^{-1}=G^{-1}+\Sigma $ and can be
formulated in terms of Grassmann variables $\psi ^{\phantom\ast }$ and $\psi
^{\ast }$. For the local Green function at a Matsubara frequency $\omega
_{\nu }=(2\nu +1)\pi /\beta $, orbital index $m$, and spin $\sigma $ one
obtains: 
\begin{equation}
G_{\nu m}^{\sigma }=-\frac{1}{{\cal Z}}\int {\cal D}[\psi ]{\cal D}[\psi
^{\ast }]\psi _{\nu m}^{\sigma \phantom\ast }\psi _{\nu m}^{\sigma \ast
}e^{{\cal A}[\psi ,\psi ^{\ast },{\cal G}^{-1}]},  \label{siam}
\end{equation}
where the single-site action ${\cal A}$ is given by
\begin{eqnarray}
\lefteqn{{\cal A}[\psi ,\psi ^{\ast },{\cal G}^{-1}]=\sum_{\nu ,\sigma ,m}\psi _{\nu
m}^{\sigma \ast }({\cal G}_{\nu m}^{\sigma })^{-1}\psi _{\nu m}^{\sigma {%
\phantom\ast }}}\nonumber \\
&&-\frac{U}{2}\!\!\!\sum_{(m\sigma )\neq (m\sigma ^{\prime})}
\!\!\int\limits_{0}^{\beta }d\tau \,\psi _{m}^{\sigma \ast }(\tau )\psi
_{m}^{\sigma \phantom\ast }(\tau )\psi _{m^{\prime }}^{\sigma
^{\prime }\ast }(\tau )\psi _{m^{\prime }}^{\sigma ^{\prime }\phantom%
\ast }(\tau ). 
\end{eqnarray}

\subsection{Quantum Monte-Carlo method (QMC)}

To solve the effective single-site problem defined above we will apply 
the QMC method
which allows for a numerically exact solution~\cite{QMC}. To this end, the
imaginary time is first discretized into $\Lambda $ steps of size~$\Delta
\tau $ 
($\Delta \tau = 0.25$~eV$^{-1}$ throughout this work), 
and $\tau _{t}:=t\Delta \tau $ with integer~$t$. Then, in a second
step, the Hubbard-Stratonovich transformation 
\begin{eqnarray}
\lefteqn{\exp \left\{ \frac{\Delta \tau U}{2}(\psi _{mt}^{\sigma \ast }\psi
_{mt}^{\sigma \phantom\ast }-\psi _{m^{\prime }t}^{\sigma ^{\prime }\ast
}\psi _{m^{\prime }t}^{\sigma ^{\prime }\phantom\ast })^{2}\right\}=}&&
\nonumber  \\ && \frac{1%
}{2}\!\!\!\sum_{s_{m\sigma ;m^{\prime}\!\sigma ^{\prime }}^{t}\!=\!\pm 1}\!\!\!\exp \left\{
\lambda s_{m\sigma ;m^{\prime }\sigma ^{\prime }}^{t}(\psi _{mt}^{\sigma
\ast }\psi _{mt}^{\sigma \phantom\ast }-\psi _{m^{\prime }t}^{\sigma
^{\prime }\ast }\psi _{m^{\prime }t}^{\sigma ^{\prime }\phantom\ast
})\right\} 
\end{eqnarray}
is employed, which replaces the interacting system by a sum of $%
2^{(2M^{2}-M)\Lambda }$ different non-interacting systems, where $M$ is the
number of interacting orbitals. Each non-interacting system can be solved by
a Gauss integration which yields a contribution $({\bf M}^{{\bf s}})^{-1}\det
({\bf M}^{{\bf s}})$ to the Green function in imaginary time, parameterized by
($t$, $t^{\prime }$), where 
\begin{eqnarray}
\lbrack {\bf M}_{m}^{\sigma {\bf s}}]_{tt^{\prime }}&=&\Delta \tau ^{2}\left[
({\bf G}_{m}^{\sigma })^{-1}+{\bf \Sigma }_{m}^{\sigma }\right]
_{tt^{\prime }}\nonumber \\ &&
-\delta _{tt^{\prime }}\sum_{m^{\prime }\sigma ^{\prime
}}\lambda _{m\sigma ;m^{\prime }\sigma ^{\prime }}\;\tilde{\sigma}_{m\sigma ;
m^{\prime }\sigma ^{\prime }}\;s_{m\sigma ;m^{\prime }\sigma
^{\prime }}^{t}.  \label{M1}
\end{eqnarray}
Here, $\tilde{\sigma}_{m\sigma ;m^{\prime }\sigma ^{\prime }}=2\Theta (\sigma
^{\prime }-\sigma +\delta _{\sigma \sigma ^{\prime }}[m^{\prime }-m])-1$
has a different sign if ($m\sigma $) and ($m^{\prime }\sigma ^{\prime}$)
are exchanged.
Since a full summation over all non-interacting systems is
computationally impossible, the Monte-Carlo method is employed for
importance sampling. Details of the one-band QMC algorithm in the context
of DMFT can be found in Refs.~\cite{QMC,georges96} and for the band-degenerate
case in Refs.~\cite{rozenberg}.

\section{Results and Discussion}

The stoichiometric compound LaTiO$_{3}$ has a perovskite crystal structure
with a small orthorhombic distortion \linebreak
($\angle~Ti-O-Ti~\approx~155^{\circ }$)~\cite{maclean79}.
Below $T_{N}=125$~K~\cite{gopel} it is an antiferromagnetic insulator~\cite{eitel86}
with a Ti magnetic moment of 0.45~$\mu_{B}$ and a small
energy gap of approximately $0.2$~eV~\cite{okimoto95,goral82}. 
At doping $x=0.05$, La$_{1-x}$Sr$_{x}$TiO$_{3}$ undergoes an
insulator-to-metal transition and becomes a correlated paramagnetic metal
with a strongly enhanced susceptibility and electronic specific heat
coefficient~\cite{tokura93}.

\begin{figure}[tbp]
 \centering
\includegraphics[clip=true,width=8cm]
 {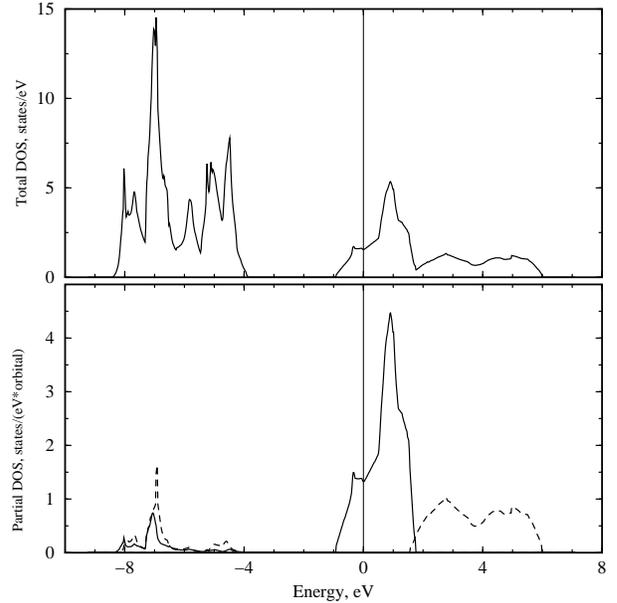}

\caption{Densities of states of LaTiO$_3$ calculated with LDA-LMTO. Upper
figure: total density of states; lower figure: partial $t_{2g}$ (solid
lines) and $e_g$ (dashed lines) densities of states.}
\label{ldados}
\end{figure}

The standard LDA calculation for undoped 
LaTiO$_3$ yields a density of states (Fig. \ref{ldados})
which is typical for early transition metal oxides with a completely filled $2p$
oxygen band ranging from $-8.2$~eV to $-4.0$~eV and a partially filled Ti-3$d$
band above it. Since the Ti-ion has an octahedral coordination of oxygen
ions, the Ti-3$d$ band is split into three degenerate $t_{2g}$ and two
degenerate $e_{g}$ subbands which do not mix. Titanium is three-valent in
LaTiO$_{3}$ and the corresponding formal ionic configuration is $d^{1}$.
This implies a partially filled $t_{2g}$ subband containing one electron
with a total capacity of six electrons. The $e_{g}$ subband is empty and
situated just above the $t_{2g}$ subband. In order to simplify our
calculations the real orthorhombic crystal structure was replaced by a cubic
structure with the same volume. This approximation leads to a slight
overestimation of the effective bandwidth.

The $t_{2g}$ band is the only partially filled band and is well separated
from other bands. Therefore in our DMFT calculation we took only Coulomb
interactions between electrons in $t_{2g}$ orbitals into account. The
resulting problem is equivalent to a Hubbard model with three degenerate
bands. The DMFT chemical potential was adjusted to yield  the doping 
$x=0.06$, i.e. 0.94 electrons in the 
$t_{2g} $ band. We employ Eq.~(\ref{ELDAalimit}) within constrained
LDA \cite{Parameters}, i.e., changing the number of $t_{2g}$-electrons only, to calculate $U$.
Our LMTO-ASA calculation (TB-LMTO-ASA code  of Andersen 
    and coworkers \cite{LMTO} version 47)
yields a fully-screened Coulomb interaction of $U=4.2$ eV  within the 
basis Ti(4s,4p,3d) La(6s,6p,5d) O(2s,2p) at a Wigner Seitz 
radius of 2.37 a.u. for Ti.
Our result has to be compared to that of Solovyev et al.~\cite{solovyev96}
who obtained $U=3.2$ eV employing the ASA-LMTO method
within orthogonal representation.
This shows that the ab-initio calculation of $U$, 
which is the interaction between particular ($t_{2g}$)
orbitals, is rather sensitive to the orthogonality of the
 wave functions and, also, to the choice of the orbitals.
Unless specified otherwise, we will, thus, use
$U=4$ eV and keep in mind that  the 
inherent uncertainty is about 0.5 eV.

\begin{figure}[tbp]
\centering
\includegraphics[clip=true,width=8cm]
 {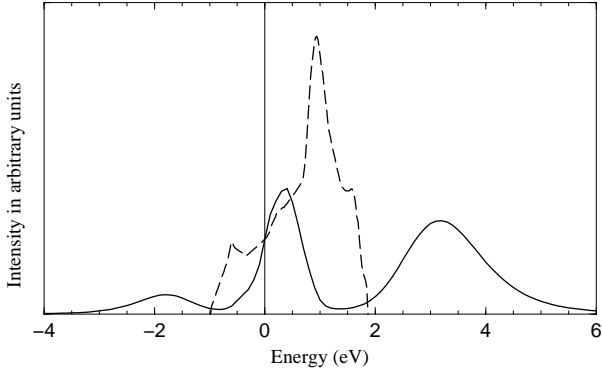}

\caption{Partial $t_{2g}$ densities of states of LaTiO$_3$ calculated with
LDA+DMFT(QMC) (solid lines) and LDA (dashed lines).}
\label{doplatio}
\end{figure}

In Fig.~\ref{doplatio}, the spectral function obtained from our \linebreak
LDA+DMFT(QMC) calculation at temperature $T\approx$~1000~K 
is compared with the non-interacting $t_{2g}$ density of states.
One can see the typical features of the spectra of strongly correlated
systems: a lower Hubbard band, a well pronounced
quasi-particle peak, and an upper Hubbard
band. While for the non-interacting case 100\% of the spectral weight is
located in the quasi-particle band, the LDA+DMFT spectra are
characterized by a spectral weight transfer from the quasiparticle band to
the Hubbard bands and a narrowing of the quasiparticle band.

\begin{figure}[tbp]
\includegraphics[clip=true,width=8cm]
 {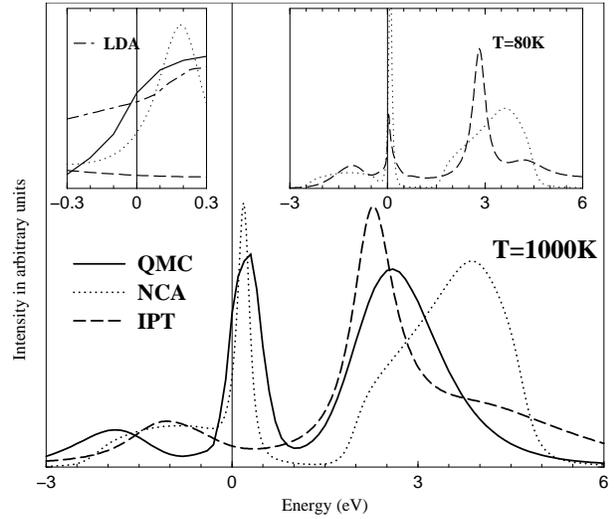}

\caption{Comparison of the spectral densities of La$_{1-x}$Sr$_{x}$TiO$_{3}$
($x=0.06$) as calculated by LDA+DMFT using
the approximations IPT and NCA, with
the numerically exact QMC-result at $T=0.1$~eV, i.e., approximately 1000~K,
and $U=4$~eV.
Inset left: Behavior at the Fermi level including the LDA DOS.
Inset right: NCA and IPT spectra for the temperature~80~K.}
\label{theory}
\end{figure}

The QMC simulations performed in this paper to solve the effective
multi-band single-impurity model provide a numerically exact solution, but
require a large computational effort which 
restricted our calculations to temperatures of the order of 
$T=0.1$~eV ($\approx$~1000~K). Since this technique
yields results only at imaginary \linebreak
(Matsubara-) frequencies  the calculation of 
the spectral function requires an analytic continuation
of the spectral function
via, e.g., the maximum entropy method~\cite{MEM}.
Previous LDA+DMFT investigations of the photoemission spectra of 
La$_{1-x}$Sr$_{x}$TiO$_{3}$ used  a variant of IPT for doped multi-band systems~\cite{poter97} and NCA~\cite{zoelfl99} to solve the DMFT-equations,
i.e., approximate techniques.
In Fig.~\ref{theory}, we compare the results obtained
within these approximations with the numerically exact QMC
simulation, all at~$T\approx 1000$~K. 
One notes that within IPT the shape of the upper Hubbard band is not correct.
Moreover, there is no quasiparticle peak at 1000~K,
the reason being
that IPT underestimates the Kondo temperature
considerably such
that the very narrow  quasiparticle peak found at low temperatures 
(see right inset of Fig.~\ref{theory})
disappears already at about 250~K. A similarly narrow IPT  
quasiparticle peak
was found in a three-band model study with Bethe-DOS by
 Kajueter and Kotliar \cite{Kajueter}.
While NCA comes off much better than IPT
it still underestimates the width of the quasiparticle peak by a factor of two. 
Furthermore, the position of the quasiparticle peak is too
close to the lower Hubbard band.
In the left inset of Fig.~\ref{theory}, the behavior at the Fermi level is shown. At the Fermi
level,
the NCA yields a spectral function which is almost by a factor  two
too small.
 The
shortcomings of the NCA-results  appear to result from the
well-known problems which this approximation scheme encounters
already in the single-impurity Anderson model at low temperatures
and/or low frequencies~\cite{MH,NCAdeficit}.
Similarly, the deficiencies of the IPT-results are not 
entirely surprising in view of
the semi-phenomenological nature of this approximation, especially for a
system off half filling. This comparison
shows that the choice of the method used to solve the DMFT
equation is indeed important.

\begin{figure}[tbp]
 \centering
\includegraphics[clip=true,width=8cm]
 {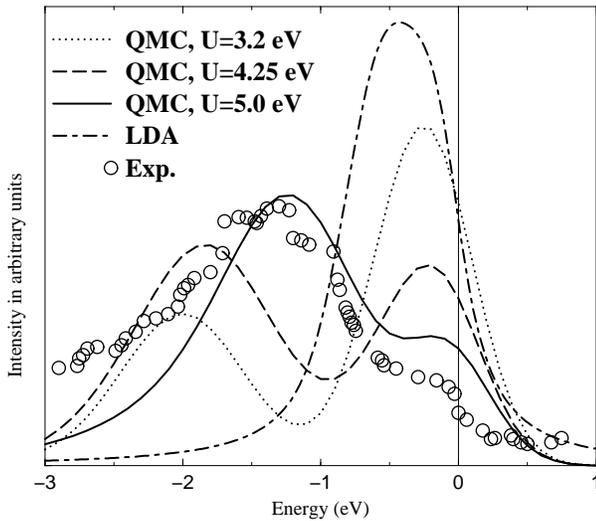}

\caption{Comparison of the experimental photoemission spectrum
\protect~\cite{fujimori}, the  LDA result, and the 
LDA+DMFT(QMC) calculation for
 LaTiO$_{3}$ with 6\% hole doping and different
Coulomb interaction $U=3.2$, $4.25$, and $5$eV.}
\label{explatio}
\end{figure}

Photoemission spectroscopy of the early transition metal oxides provides a
direct tool for the study of the electronic structure of strongly correlated
materials. A comparison of the experimental photoemission 
spectra~\cite{fujimori}
with results obtained from LDA and LDA+DMFT(QMC) at 1000~K~\cite{Note}
are shown in Fig.~\ref{explatio}. To take into account the uncertainty in 
$U$,  we present results
for $U=3.2$, $4.25$ and $5$ eV. All spectra are 
multiplied with the Fermi step function 
and  Gaussian-broadened  with a broadening parameter
of 0.3~eV
  to simulate the experimental
 resolution~\cite{fujimori}.
The
LDA band structure calculation clearly fails to reproduce the broad band
observed in the experiment at 1-2~eV below the Fermi energy~\cite{fujimori}. 
Taking
the correlations between the electrons into account, this lower
band is easily
identified as the lower Hubbard band whose spectral
weight originates from  the quasiparticle 
band at the Fermi energy and increases with $U$.
The best  agreement with experiment concerning 
the relative intensities of the Hubbard band and the 
quasi-particle peak and, also, the
position of the Hubbard band is found for $U=5$ eV
\cite{noteU5,IPTNCA}.
The value  $U=5$ eV
is still compatible  with the ab-initio calculation of this parameter.
One should also note that the photoemission
experiment is sensitive to
surface properties. Due to the reduced coordination number
at the surface, the bandwidth is likely to be smaller 
and the  Coulomb interaction to be less screened, i.e., larger.
Both effects make the system more correlated and, thus,
might also explain why better agreement is
found for $U=5$ eV.
Besides, the polycrystalline nature of the sample
and, also, spin and orbital \cite{Keimer}  fluctuation, not taken into
account in the LDA+DMFT approach, could further reduce the 
quasiparticle weight.

In conclusion, the LDA+DMFT(QMC) approach is shown to be
 a workable computational
scheme which merges the conventional band structure approach with a recently
developed many-body technique in combination with a numerically reliable
evaluation method. Thereby, it provides a powerful tool for future
{\em ab initio} investigations of real materials with strong electronic
correlations. The LDA+DMFT(QMC) approach not only
explains the existence of the lower Hubbard band in
doped LaTiO$_{3}$, but also, in contrast to LDA,
reproduces the qualitative picture of the spectral weight transfer from
the quasiparticle band to the lower Hubbard band, the position of the
lower Hubbard band, and 
the narrowing of the quasiparticle band.

\section{Acknowledgment}
We are grateful to R. Claessen, J. L\ae gsgaard, 
Th.~Pruschke, G. A. Sawatzky, J. Schmalian, and
M. Z\"olfl for useful discussions.
This work was supported in part by the Sonderforschungsbereich 
484 of the Deutsche Forschungsgemeinschaft 
and the Russian Foundation for Basic Research (RFFI-98-02-17275).


\begin{thebibliography}{99}
\bibitem{NCAdeficit}  NCA is a self-consistent perturbation theory for the
single-impurity Anderson model and is often a fast method to obtain results
for this model and also within DMFT. However, NCA is known to violate Fermi
liquid properties at temperatures much lower than the smallest energy scale
of the problem and whenever charge excitations become important~\cite{MH,Pruschke93}.
Hence, in some parameter ranges it fails to account for the
proper low-energy physics and must therefore be applied with considerable
care~\cite{Pruschke93}.
\bibitem{Note}
At present, QMC simulations of the DMFT equations
 are not feasible at the experimental temperature
(80K). We note, however, that no 
intrinsic temperature dependence was observed in the experiment~\cite{fujimori}, 
at least up to room temperature.
\bibitem{noteU5}
In Fig.~\ref{explatio}, the distance between
the lower Hubbard band and the chemical potential
is seen to decrease with increasing $U$. 
Note, however, that the 
upper Hubbard band is also shifted to higher energies.
\bibitem{IPTNCA}
If one multiplies the IPT and NCA results of Fig. 3 with the Fermi function
and employs a broadening of 0.3 eV one obtains a spectrum where the 
lower Hubbard band is nearer to the Fermi energy and 
which has
a less pronounced minimum or shoulder  between 
 Hubbard and quasi-particle band.
\end{thebibliography}
\end{document}